\documentclass{appolb}
\usepackage{graphicx}
\begin{document}
\title{Origin of the Orbital Ordering in LaMnO$_3$%
\thanks{Presented at the Strongly Correlated Electron Systems 
Conference, Krak\'ow 2002}%
} 

% Authors and Affiliations

\author{  Olga Sikora and Andrzej M. Ole\'s
\address{ Marian Smoluchowski Institute of Physics, Jagellonian 
          University,\\
          Reymonta 4, PL-30059 Krak\'ow, Poland}
\date{10 July 2002}
}
\maketitle

% Abstract

\begin{abstract}
We use the temperature of the structural phase transition to determine 
the Jahn-Teller (JT) coupling constant in the model derived for 
LaMnO$_3$ which includes both the superexchange between $S=2$ spins and 
the JT effect. We also investigate the dependence of the exchange 
constants on the value of the on-site Coulomb element $U$, and on the 
orbital ordering.\\
Journal ref.: O. Sikora and A. M. Ole\'s, 
              Acta Phys. Pol. B {\bf 34}, 861 (2003).

\end{abstract}

\PACS{75.30.Et, 75.47.Lx, 75.50.Ee}

%%%%%%%%%%%%%%%%%%%%%%%%%%%%%%%%%%%%%%%%%%%%%%%%%%%%%%%%%%%%%%%%%%%%%%%%
%%                       SE and JT in eg systems
%%%%%%%%%%%%%%%%%%%%%%%%%%%%%%%%%%%%%%%%%%%%%%%%%%%%%%%%%%%%%%%%%%%%%%%%
For degenerate $e_g$ orbitals the superexchange (SE) interactions 
involve both spin and orbital degrees of freedom. Therefore, these 
interactions alone could stabilize the orbital ordering, either in 
cuprates \cite{Ole00}, or in manganites \cite{Fei99}. However, large 
lattice distortions observed in LaMnO$_3$ suggest that the Jahn-Teller 
(JT) effect also plays an important role in inducing the staggered 
orbital ordering. In particular, if the SE interactions are ignored, the 
observed temperature of the structural transition $T_s=780$ K leads to 
a large JT coupling constant \cite{Mil96}. We show that the orbital 
interactions which originate either from the SE or from the JT effect 
support each other, so one can deduce the JT coupling constant only by 
analyzing all the terms responsible for the orbital ordering 
\cite{Oki02}.

%%%%%%%%%%%%%%%%%%%%%%%%%%%%%%%%%%%%%%%%%%%%%%%%%%%%%%%%%%%%%%%%%%%%%%%%
%%                                 model
%%%%%%%%%%%%%%%%%%%%%%%%%%%%%%%%%%%%%%%%%%%%%%%%%%%%%%%%%%%%%%%%%%%%%%%%
We consider a model Hamiltonian for LaMnO$_3$ \cite{Fei99}:
\begin{equation} 
H=H_{e}+H_{t}+H_{\rm JT},
\end{equation}
which includes the SE due to $e_g$ ($H_e$) and $t_{2g}$ ($H_t$) 
electrons, and the JT term ($H_{\rm JT}$). The term $H_{e}$ originates 
from the virtual hopping to four $t_{2g}^3e_g^2$ excited states:
one high-spin $^6\!\!A_1$ state and three low-spin $^4\!\!A_1$, $^4\!E$, 
$^4\!\!A_2$ states, with the excitation energies:
$\varepsilon(^6\!\!A_1)=U-3J_{\rm H}$,
$\varepsilon(^4\!\!A_1)=U+2J_{\rm H}$,
$\varepsilon(^4\!E\,)=U+\frac {8}{3}J_{\rm H}$,
$\varepsilon(^4\!\!A_2)=U+\frac{16}{3}J_{\rm H}$ \cite{Fei99}. The 
Coulomb and exchange elements are: $U\simeq 5.9$ eV and $J_{\rm H}=0.69$ 
eV \cite{Miz96}. One finds
\begin{eqnarray}
\label{egterm}
H_{e}=\frac{1}{16}\sum_{\langle ij\rangle}\left\{
 - \frac{8}{5} \frac{t^2}{\varepsilon(^6\!\!A_1)}
   \left(\vec{S}_i\cdot\vec{S}_j+6\right)
   {\cal P}_{\langle ij\rangle}^{\zeta\xi}\right.+              
    \left(\vec{S}_i\cdot\vec{S}_j-4\right) \nonumber \\
\times \left. \left[ \left(\frac{t^2}{\varepsilon(^4\!E)}
   + \frac{3}{5}\frac{t^2}{\varepsilon(^4\!\!A_1)} \right)
   {\cal P}_{\langle ij\rangle}^{\zeta\xi}            
+ \left( \frac{t^2}{\varepsilon(^4\!E)}
   + \frac{t^2}{\varepsilon(^4\!\!A_2)} \right)
   {\cal P}_{\langle ij\rangle}^{\zeta\zeta} \right] \right\},
\end{eqnarray}
for the SE between $S=2$ spins, where $t$ is the hopping element between 
two directional $3z^2-r^2$ orbitals along the $c$-axis (here we take 
$t=0.48$ eV which is consistent with the experimental exchange constants 
\cite{Mou99}), and the operators
${\cal P}_{\langle ij\rangle}^{\alpha\beta}=
 P_i^{\alpha}P_j^{\beta}+P_i^{\beta}P_j^{\alpha}$ consist of the orbital 
projections at both sites, $P_i^{\alpha}$ and $P_j^{\beta}$ 
\cite{Fei99}. Virtual hopping of $t_{2g}$ electrons gives the second SE 
contribution, 
$H_{t}=\frac{1}{4}J_t\sum_{\langle ij\rangle}
                    (\vec{S}_i\cdot\vec{S}_j-4)$,
with an average exchange constant $J_t\simeq 2.9$~meV.
The JT term induced by the oxygen distortions \cite{Mil96}:
\begin{equation}
H_{\rm JT}=\kappa\sum_{\langle ij\rangle}\left(
           {\cal P}_{\langle ij\rangle}^{\zeta\zeta}
         -2{\cal P}_{\langle ij\rangle}^{\zeta\xi  }
          +{\cal P}_{\langle ij\rangle}^{\xi  \xi  }\right),
\end{equation}
favors alternating ordering of orbitals parallel ($|\zeta\rangle$) and 
perpendicular ($|\xi\rangle$) to the bond direction, such as 
$|z\rangle=3z^2-r^2$ and $|x\rangle=x^2-y^2$ along $c$-axis.

%%%%%%%%%%%%%%%%%%%%%%%%%%%%%%%%%%%%%%%%%%%%%%%%%%%%%%%%%%%%%%%%%%%%%%%%
%%                          orbital ordering  
%%%%%%%%%%%%%%%%%%%%%%%%%%%%%%%%%%%%%%%%%%%%%%%%%%%%%%%%%%%%%%%%%%%%%%%%
We assume the $A$-AF spin structure [spins parallel in $(a,b)$ plane and 
staggered along $c$ axis], and $C$-type orbital ordering, with the 
orbital state
\begin{equation}
\label{theta} 
|\theta\rangle = \cos\frac{\theta}{2}|z\rangle 
               + \sin\frac{\theta}{2}|x\rangle,
\end{equation}
alternating ($|\pm\theta\rangle$) in $(a,b)$ plane and repeating itself 
along $c$ direction. By averaging the orbital operators  one can derive 
an effective Heisenberg model with intraplanar ($J_{ab}$) and interplanar
($J_c$) exchange constants, which give the N\'eel temperature $T_{\rm N}$ 
in the mean-field approximation (MFA).

\begin{figure}[!ht]
\begin{center}
\includegraphics[width=\textwidth]{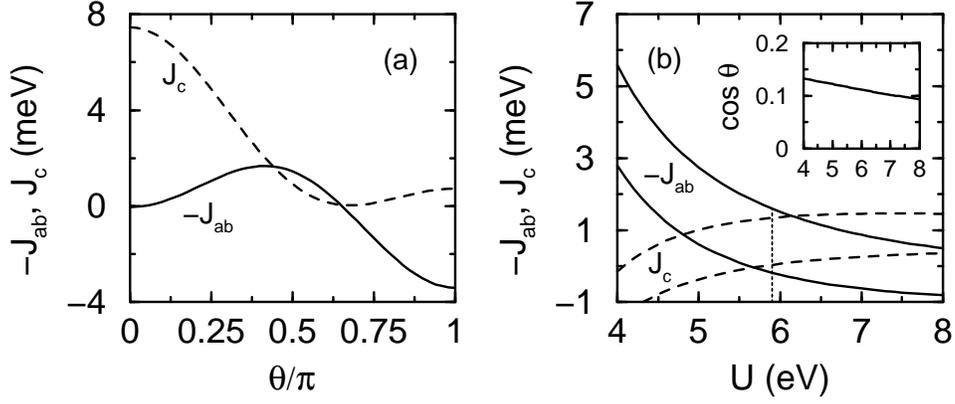}
\end{center}
\caption{ 
Exchange constants $J_{ab}$ (solid lines) and $J_c$ (dashed lines) as 
functions of:
(a) orbital angle $\theta/\pi$ for $U=5.9$ eV, 
(b) $U$ for the optimal orbital angle $\theta_{opt}$ (higher lines)
    shown in the inset, and for a fixed ordering with $\theta=2\pi/3$
    (lower lines).}
\label{jabc}
\end{figure}

\begin{table}[h]
\label{angles}
\caption{Exchange constants $J_{ab}$ and $J_{c}$ and $e_g$ contributions
$J_{ab}(e_g)$ and $J_{c}(e_g)$ (all in meV) obtained with $U=5.9$ eV,
compared with experimental values.}
\begin{center} \begin{tabular}{|c|c|c|c|c|c|}
\hline
$\theta$ & orbital state & $J_{ab}(e_g)$ & $J_{c}(e_g)$ 
                              & $J_{ab}$    & $J_{c}$         \\
\hline
$\pi/3$  &  $|y^2-z^2\rangle/|x^2-z^2\rangle$  & -2.23 & 2.67 & -1.50 & 3.40 \\
$\pi/2$  & $(|x\rangle+|z\rangle)/
           (|x\rangle-|z\rangle)\rangle$       & -2.18 & 0.20 & -1.45 & 0.92 \\
$2\pi/3$ & $|3x^2-r^2\rangle/|3y^2-r^2\rangle$ & -0.54 & -0.69 & 0.18 & 0.03 \\
$\sim 83^{\circ}$ & $|+\theta_{opt}\rangle/
                     |-\theta_{opt}\rangle$    & -2.33 & 0.62 & -1.60 & 1.34 \\
  -      & exp \cite{Mou99} & - & - & -1.66 & 1.16 \\
\hline
\end{tabular} \end{center}
\end{table}

%%%%%%%%%%%%%%%%%%%%%%%%%%%%%%%%%%%%%%%%%%%%%%%%%%%%%%%%%%%%%%%%%%%%%%%%
%%                         exchange constants
%%%%%%%%%%%%%%%%%%%%%%%%%%%%%%%%%%%%%%%%%%%%%%%%%%%%%%%%%%%%%%%%%%%%%%%%
The exchange constants $J_{ab}$ and $J_c$ depend strongly on the orbital
ordering (\ref{theta}), and the $A$-AF phase is stable for 
$\theta/\pi<0.65$, while $G$-AF order occurs for $\theta/\pi>0.65$ [Fig. 
\ref{jabc}(a)]. The values obtained for the optimal orbital state 
$|\theta_{opt}\rangle$ which gives the lowest energy at $T=0$ agree well 
with the experimental data \cite{Mou99} (Table 1). From the dependence
of $J_{ab}$ and $J_c$ on the value of $U$ we conclude that only values
of $U\leq 6$ eV are able to explain the experimental result 
$|J_{ab}|>J_c$ [Fig. \ref{jabc}(b)]. Moreover, $J_{ab}$ and $J_c$ 
obtained for $\theta=2\pi/3$, often assumed for LaMnO$_3$, are never 
close to experiment.   

\begin{figure}[!ht]
\begin{center}
\includegraphics[width=\textwidth]{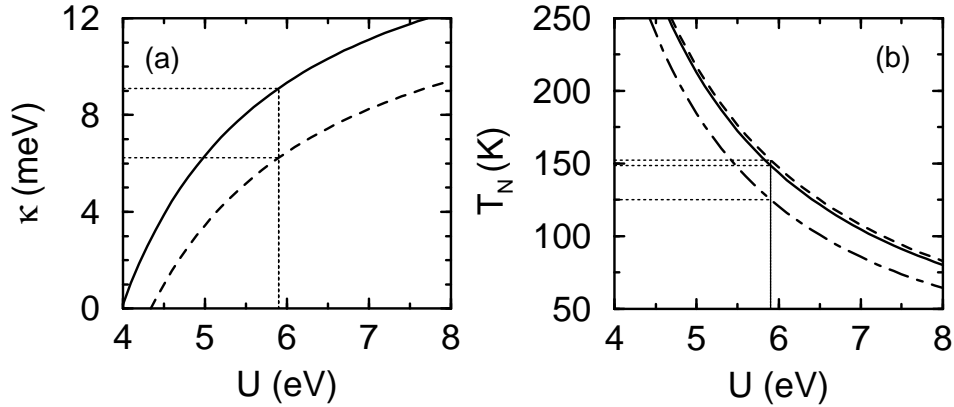}
\end{center}
\caption{
Values of: 
(a) JT interaction $\kappa$, and 
(b) the N\'eel temperature $T_{\rm N}$ (obtained by reducing the value
    found in the MFA by a factor $f_{\rm N}=0.705$), 
for increasing $U$, and for $f_s=0.629$ (solid lines) and $f_s'=0.75$ 
(dashed lines). The values of $T_{\rm N}$ for the orbital ordering  
given by $\theta=\pi/2$ are shown in (b) by dot-dashed lines. }
\label{kappa}
\end{figure}

%%%%%%%%%%%%%%%%%%%%%%%%%%%%%%%%%%%%%%%%%%%%%%%%%%%%%%%%%%%%%%%%%%%%%%%%
%%                               $\kappa$
%%%%%%%%%%%%%%%%%%%%%%%%%%%%%%%%%%%%%%%%%%%%%%%%%%%%%%%%%%%%%%%%%%%%%%%%
Now we turn to the structural transition. Using realistic parameters one 
finds that $T_s \simeq 380$~K, calculated in the MFA with only the SE 
terms considered, is much lower than observed (the MFA value was reduced 
by a factor $f_s=0.629$ adequate for pseudospins 1/2).
This proves that the JT term (3)
is of importance. Considering the full Hamiltonian (1) we obtained the 
JT coupling constant $\kappa\simeq 9.1$~meV. We verified that $12\kappa$ 
corresponds to $g_{\rm JT}^2/K$ in the JT Hamiltonian derived by Okamoto 
{\it et al.\/} \cite{Oki02}. This allows us to determine 
$E_{\rm JT}\simeq 320$ meV which turns out to be significantly higher 
than the value $50<E_{\rm JT}<100$ meV estimated before. Note 
however that a different reduction factor $f_s'=0.75$ used in Ref. 
\cite{Oki02} would result in a somewhat lower value 
$E_{\rm JT}\simeq 270$ meV also in our model [Fig. \ref{kappa}(a)].

%%%%%%%%%%%%%%%%%%%%%%%%%%%%%%%%%%%%%%%%%%%%%%%%%%%%%%%%%%%%%%%%%%%%%%%%
%%                           dependence on U
%%%%%%%%%%%%%%%%%%%%%%%%%%%%%%%%%%%%%%%%%%%%%%%%%%%%%%%%%%%%%%%%%%%%%%%%
The values of $\kappa$ depend strongly on $U$ as the energies of
excitated states which contribute to the SE (2) become higher, and the 
JT term has to compensate for the reduced SE contribution [see Fig.
\ref{kappa}(a)]. Although we did not perform a self-consistent 
calculation of the orbital ordering as a function of $T$, we emphasize
that the values of $T_{\rm N}$ obtained for the 
orbital ordering given either by $\theta_{opt}$ (at $T=0$), or by 
$\theta=\pi/2$ (at $T=T_{\rm N}$), agree well with the experimental 
$T_{\rm N}^{exp}=136$ K \cite{Mou99} for $U\simeq 5.9$ eV 
[Fig. \ref{kappa}(b)].

%%%%%%%%%%%%%%%%%%%%%%%%%%%%%%%%%%%%%%%%%%%%%%%%%%%%%%%%%%%%%%%%%%%%%%%%
%%                           main conclusions
%%%%%%%%%%%%%%%%%%%%%%%%%%%%%%%%%%%%%%%%%%%%%%%%%%%%%%%%%%%%%%%%%%%%%%%%
Summarizing, we have found that the SE and the JT terms 
{\it contribute about equally\/} to the structural transition, and the 
JT coupling constant in LaMnO$_3$, $E_{\rm JT}\simeq 300$ meV, is 
at least three times larger 
than found using a simplified model of the SE \cite{Oki02}. In addition, 
we argue that the orbital state in LaMnO$_3$ has to be 
{\it significantly different\/} from $\theta=2\pi/3$, an angle which 
would give almost vanishing and AF exchange constants in all directions.

%%%%%%%%%%%%%%%%%%%%%%%%%%%%%%%%%%%%%%%%%%%%%%%%%%%%%%%%%%%%%%%%%%%%%%%%
%%
%%                          ACKNOWLEDGMENTS
%%
%%%%%%%%%%%%%%%%%%%%%%%%%%%%%%%%%%%%%%%%%%%%%%%%%%%%%%%%%%%%%%%%%%%%%%%%
This work was supported by the Polish State Committee of Scientific 
Research (KBN), Project No. 5~P03B~055~20.

%%%%%%%%%%%%%%%%%%%%%%%%%%%%%%%%%%%%%%%%%%%%%%%%%%%%%%%%%%%%%%%%%%%%%%%%
%%
%%                            REFERENCES
%%
%%%%%%%%%%%%%%%%%%%%%%%%%%%%%%%%%%%%%%%%%%%%%%%%%%%%%%%%%%%%%%%%%%%%%%%%


\begin{thebibliography}{}

\bibitem{Ole00} A. M. Ole\'s, L. F. Feiner, and J. Zaanen,
 \textit{Phys. Rev.} \textbf{B61}, 6257 (2000).

\bibitem{Fei99} L. F. Feiner and A. M. Ole\'s,
 \textit{Phys. Rev.} \textbf{B59}, 3295 (1999).

\bibitem{Mil96} A. J. Millis,
 \textit{Phys. Rev.} \textbf{B53}, 8434 (1996).

\bibitem{Oki02} S. Okamoto, S. Ishihara and S. Maekawa,
 \textit{Phys. Rev.} \textbf{B65}, 144403 (2002).

\bibitem{Miz96} T. Mizokawa and A. Fujimori, 
 \textit{Phys. Rev.} \textbf{B54}, 5368 (1996).

\bibitem{Mou99} F. Moussa, M. Hennion, G. Biotteau, and
J. Rodr\'{\i}guez-Carvajal \textit{Phys. Rev.} \textbf{B60}, 12299 (1999).

\end{thebibliography}
\end{document}